\documentstyle[11pt]{article}
\begin{document}
\begin{flushright}SJSU/TP-98-15\\January 1998\end{flushright}
\vspace{1.7in}
\begin{center}\Large{\bf Does Schr\"{o}dinger's Cat Know Something \\
          That Schr\"{o}dinger Does Not Know?}\\
\vspace{1cm}
\normalsize\ J. Finkelstein\footnote[1]{
        Participating Guest, Lawrence Berkeley National Laboratory\\
        \hspace*{\parindent}\hspace*{1em}
        e-mail: JLFinkelstein@lbl.gov}\\
        Department of Physics\\
        San Jos\'{e} State University\\San Jos\'{e}, CA 95192, U.S.A
\end{center}
\begin{abstract}
   Macroscopic objects { \em appear} to have definite positions.  In a
   many-worlds interpretation of quantum theory, this appearance is an
   illusion; the correct view is the ``view from outside'' in which even
   macroscopic objects are in general in a superposition of different
   positions. In the Bohm model, objects {\em really are} in definite
   positions.  This additional aspect of reality is accessible only from
   the ``inside''; thus in the Bohm model the view from inside can be 
   more correct than is the view from outside. 
\end{abstract}
\newpage
Quantum mechanics is an enormously successful theory.  There has been
experimental verification of quantum predictions for processes involving
distances ranging from about $10^{-16}$ cm (photoproduction in \={p}p
scattering, ref.\ \cite{CDF}) to about 10 km (Bell's inequality for
entangled photons, ref\ \cite{GSN}).  The successes of quantum mechanics 
come in spite of the fact that the most
straightforward interpretation of one of the most fundamental aspects
of quantum theory---the linear evolution implied by the  Schr\"{o}dinger
equation---would lead one to expect the existence of superpositions of
macroscopically-distinguishable states, in complete contradiction to the
way the world appears to us to be.  This discrepancy between naive
prediction and apparent observation can be ``resolved'' by postulating
that the state-vector experiences ``collapse'' as well as  Schr\"{o}dinger
evolution.  However, in this paper I will discuss other interpretations
of quantum theory, interpretations in which there is no such collapse,
and in which the state-vector does indeed evolve exactly as dictated by the
Schr\"{o}dinger equation.

The seeming implausibility of the existence of superpositions of\linebreak 
macroscopically-distinguishable states is vividly illustrated by the
story of  Schr\"{o}dinger's cat \cite{SC}. As the title of this paper implies, 
I want to discuss what someone in the position of the cat in that story might
know, and for this purpose I will suppose that a cat can be aware of
himself and of his surroundings, just as a person might be.  Any reader
unable to imagine such a sentient cat can just substitute the word 
``human'' for ``cat'' wherever it appears in this discussion. However, since
I doubt that my reader's credulity would extend to the case of a sentient
{\em dead} cat, I will alter the story somewhat, so that the two possibilities
for the cat correspond to being in two different positions, rather than to
life or death.

So here is the story which I wish to discuss.  An experimenter, whom I will
identify as being  Schr\"{o}dinger himself, places into a box a cat, together
with a radioactive sample, such that perhaps in the course of an hour one
of the atoms decays, but also, with equal probability, perhaps none do.
The cat has been trained so that, at the end of the hour, if there has been 
a decay he moves to the left side of the box, but if there has not been any
decay he moves to the right side. After placing the cat and the sample in
the box,  Schr\"{o}dinger seals the box and never does look inside again,
so that the contents of the box  constitute 
an isolated system once the box is closed.

Let $B$ denote the system consisting of the contents of the box (the cat and
the radioactive atoms), and let $|\psi \rangle$ be the state-vector for
that system.  For the time $t_{\mbox{initial}}$ when the box is first 
closed, we can write
\begin{equation}
 |\psi(t_{\mbox{initial}})\rangle = |\mbox{atoms, initial}\rangle
                            |\mbox{cat, initial}\rangle .
\end {equation}         
I wish to discuss the situation within the box at a time after the hour
has passed, which I will refer to simply as time $t$. I am supposing that
the state-vector of a system, even one containing macroscopic and/or
sentient components, never experiences collapse, so I write
\begin{equation}
 |\psi (t)\rangle = \frac{1}{\sqrt{2}}(|\mbox{atom decayed}\rangle
 |\mbox{cat at left}\rangle + |\mbox{no atom decayed}\rangle
                              |\mbox{cat at right}\rangle ) .
\end{equation}
We could if we wished write a state-vector for the combined system consisting
of $B$ together with  Schr\"{o}dinger, but that state-vector would simply
be the product of $|\psi(t)\rangle $ as given in eq.\ (2) and the 
state-vector for  Schr\"{o}dinger.  

Now, at time $t$ it will certainly
appear to the cat that it is in a definite position (just as it appears
to each of us that we are in a definite position),  although the state-vector
in eq.\ (2) indicates a superposition of different positions.
As suggested by its title, this paper is devoted to a discussion, using the
example of  Schr\"{o}dinger's cat, of the following question:
Does  the appearance of a definite position correspond to some truth 
(as opposed to being an illusion), and if so,
is this a truth which could also be known by  an outside observer?
I will examine the answers to this question in two versions of no-collapse
interpretations of quantum mechanics.  One version is the many-worlds
interpretation \cite{Ev} in which the state-vector is taken to represent
a complete description of a system.  The other version assumes that the
state-vector description is {\em not} complete; I will discuss in
particular the interpretation given by de Broglie \cite{dB} and by Bohm
\cite{Bohm} (which I will call simply the Bohm model),
but similar considerations would
presumably apply to, e.g., the modal interpretations of refs.\ \cite{modal}.

\vspace{2.5ex}
I shall take the many-worlds interpretation of quantum mechanics \cite{Ev} 
to be  based on the following two assertions: 
\begin{itemize}
\item {\bf Completeness:} The state of an isolated system is completely
     described by a Hilbert-space vector (which I will denote by 
     $|\psi\rangle$).
\item {\bf No collapse:} The state-vector $|\psi\rangle $ evolves according
    to the  Schr\"{o}dinger equation
    \begin{equation}
       H |\psi\rangle = i\hbar \frac{\partial}{\partial t} |\psi\rangle .
    \end{equation}
\end{itemize}
Since in our story  Schr\"{o}dinger has prepared the initial state of the
system $B$, we may assume that he know its initial state-vector, as given
in eq.\ (1).  Then if he knows the rate of decay of the atoms and the 
way the cat has been trained (i.\ e., if he knows the Hamiltonian for $B$)
he can use eq.\ (3) to  calculate $|\psi(t)\rangle$ for the later
time $t$.  Thus we may assume that  Schr\"{o}dinger knows the state vector
$|\psi(t)\rangle$, even though he does not interact with the system $B$ after
the initial time.  

We can also imagine that the cat knows the state vector $|\psi(t)\rangle$, for
it too might know how the system was prepared and might be able to solve
eq.\ (3).  That would be an example of an inhabitant of one ``branch'' of
the many-world state-vector being aware of the existence of another branch, 
but that is certainly allowed in the many-worlds 
interpretation.\footnote[2]{Vaidman \cite{V} fancifully describes a
        {\em single particle} which inhabits one branch and yet is aware
        of the existence of another branch.}
For a more realistic example, consider an experimenter who prepares an
electron in an eigenstate of $S_x$ and then measures
its value of $S_z$ to be $+\frac{1}{2}\hbar$. Since she remembers how
the spin was originally prepared, she knows (if she is an adherent of the
many-worlds interpretation) that there is another branch in which $S_z$
was measured to be $-\frac{1}{2}\hbar$. 

However, the fact that the cat might be thought to know {\em as much as}
does  Schr\"{o}dinger does not address the question which forms the
title of this paper, which is whether the cat can know something which
Schr\"{o}dinger does not. Although a great deal can be said about
what the cat might know (see for example ref.\ \cite{Al}), it is not
necessary to address that issue to answer our question within the
many-worlds context, for the following reason:

We have seen that  Schr\"{o}dinger, even without interacting with the
system $B$ after its initial preparation, can know the state-vector
$|\psi(t)\rangle$, and in the many-worlds interpretation there is nothing
else to know.  Thus  Schr\"{o}dinger knows everything (by which I mean
everything about $B$, of course), and therefore there is nothing else
for the cat to know.  Conversely, any interpretation in which there is
anything beyond the state-vector for the cat to know necessarily violates
the completeness assertion which I have taken to be part of the definition
of the many-worlds interpretation.  Thus, without even examining the issue
of what the cat might know, we see that the many-worlds interpretation
gives a negative answer to the question posed in the title; the cat does
{\em not} know anything that  Schr\"{o}dinger does not know.

The Bohm model \cite{dB,Bohm} subscribes to the ``no collapse'' but not
the ``completeness'' assertion of the many-worlds interpretation.  Instead,
it is asserted that each particle has,
at each time, a definite value of a position variable; I will refer to this
as the ``Bohm position'' of the particle (and by the Bohm
position of an extended object I will mean the center-of-mass of the
Bohm positions of the
object's constituent particles).  It is also generally assumed that for
macroscopic objects the observed position coincides with the Bohm position.
Logically, there is another option: to formally ascribe a Bohm position
to each particle, but when it comes to interpreting the formalism, to
completely ignore the Bohm positions and to proceed as one would in
a many-worlds interpretation.  Since neither I nor anyone that I know of
favors this option, I will not pursue it further, and will thus take the
Bohm model to include the specification that, for macroscopic objects, 
the observed position is indeed the Bohm position.\footnote[3]{As pointed
  out by Englert {\em et al.} \cite{Eng}
  for microscopic objects there are situations
  in which the Bohm position does not correspond to what one might have
  expected the position to be. For macroscopic objects this situation does
  not arise.}
For definiteness, let me say that the Bohm position of the cat in our story,
is, at time $t$, at the left.  Then we can say that the cat {\em really is}
at the left, and that the cat knows (correctly) that he is at the left.
In saying this, I am certainly not attempting to construct any theory of
consciousness, nor to say how it is that we (or a cat) come to know 
anything. I am merely combining the obvious remark that we do, after all,
perceive ourselves to be in definite locations with the feature of
the Bohm model that this perception is considered to be correct.

One might make the following objection to this:  if we say that the cat 
represented by the first term on the right-hand side of eq.\ (2)
(correctly) sees himself as being at the left, should we not also say that the
other cat in the second term (incorrectly) sees himself at the right? This
objection has no force within the Bohm model. There is no ``other'' cat;
the cat really {\em is} at the left, even though there are two terms in the
state-vector. One also might worry that, since Bohm positions can
influence each other non-locally, the Bohm model might provide some way
for  Schr\"{o}dinger to learn that the cat is at the left without his
having to look in the box.  But of course this does not happen:  since
Schr\"{o}dinger and the system $B$ are in a non-entangled state
(i.e., the joint state-vector is a product), the Bohm position of any part
of $B$ is independent of the Bohm position of any part of  Schr\"{o}dinger.
In fact, if  Schr\"{o}dinger were to look inside the box, it would not be
to learn what $|\psi(t)\rangle$ is (he already knows that), but rather to
entangle himself with $B$ so that he might know the Bohm position of the cat.
So, since in our story  Schr\"{o}dinger does not look inside the box, and
therefore does not become entangled with $B$, we can conclude that
Schr\"{o}dinger does not know that the cat is at the left.  But the cat
does know.  So the Bohm model allows a positive answer to the question
posed in the title: the cat {\em does} know something that  Schr\"{o}dinger
does not know.

\vspace{2.5ex}
We have seen that the two interpretations I have discussed give different
answers to our title question.  In the many-worlds interpretation the cat
could perhaps know as much as does  Schr\"{o}dinger, but certainly not
more, since  Schr\"{o}dinger knows everything; in the Bohm model there
is something to be known beyond what  Schr\"{o}dinger knows, and that
something can indeed be known by the cat.  Let me conclude with the
following three remarks:

\underline{First remark.} I have certainly not attempted to provide any
explanation for human (or feline!) consciousness.  The cat in the story
could have been replaced by a device which moves left or right and which
contains an internal pointer which indicates the position of the device.
Then the term $|\mbox{cat at left}\rangle$ in eq.\ (2) could be replaced by
$|\mbox{device at left, pointer points left}\rangle$, with an analogous
replacement for $|\mbox{cat at right}\rangle$. What makes the
many-worlds interpretation viable is that the resulting $|\psi(t)\rangle$
would then show a correlation between the position of the device and the
indication of the pointer.  One might perhaps maintain that, in the
context of a many-worlds interpretation, this correlation itself allows
us to say that the pointer ``knows'' the location of the device (see for
example ref.\ \cite{Al}), but in any case since the state-vector
$|\psi(t)\rangle$ and hence this correlation are known to  Schr\"{o}dinger,
we would still not say that the pointer ``knows'' something that
Schr\"{o}dinger does not.  In the Bohm model, the device really would be
at the left, and the pointer really would point left, but  Schr\"{o}dinger
would not know that.

\underline{Second remark.} Here is a different version of the story I have
been discussing.  This morning I flipped a quantum coin, to determine 
whether I should work in my office this afternoon, or whether I should go
outside for a walk instead.  Therefore, my wave function this afternoon
has support both in my office and on the walking paths 
outside.\footnote[4]{More accurately, support of the wave function of the 
entangled system consisting of me and the coin includes values of my
position variable both inside and outside.} In fact, here I am working,
and so I know that my Bohm position is here in the office.  My colleague,
who is familiar with my practice of consulting that coin but has not seen 
me today, does perhaps know my wave function, but does not know my 
Bohm position.

Is this at all surprising?  In the Bohm model, the situation is simply
that I see where I am, but my colleague does not see me, so {\em of course}
I know something that my colleague does not know.  What perhaps is
surprising is that, in the many-worlds interpretation, I {\em don't}
know anything unknown to my colleague, since there are no ``facts''
beyond those represented in the wave function.

\underline{Last remark.} Returning now to the first version of the story
with the cat, we have said that, since according to eq.\ (3) the 
state-vector evolves deterministically, knowledge of the state-vector
of system $B$ at the initial time is sufficient for knowledge at the
later time $t$.  This is in contrast to interpretations in which the
state-vector can suffer collapse; in those interpretations, knowledge of the
initial state-vector is not sufficient, so that an additional measurement
would be required.  Of course  Schr\"{o}dinger did have to interact with
the system $B$ in order to prepare it in its initial state, but after that
initial interaction, he can simply calculate $|\psi\rangle$ at any later
time.  So $|\psi(t)\rangle$ might be said to represent ``public'' 
reality which is accessible to anyone who knows its initial value.
And in the many-worlds interpretation, that is all the reality there is.
In the Bohm model, there is an additional aspect to reality, namely the
Bohm positions.  The Bohm model is also deterministic, but  Schr\"{o}dinger
does not know the Bohm position of the cat at time $t$ because (although
he did prepare the initial state of $B$) he does not know well enough the
initial Bohm positions. So this additional aspect of reality is only
accessible from within the system $B$. 
   
\vspace{1cm}
Acknowledgement:
I would like to acknowledge the hospitality of the
Lawrence Berkeley National Laboratory.


\newpage
\begin{thebibliography}{99}
\bibitem{CDF} CDF collaboration: F.\ Abe {\it et al.}, Phys.\ Rev.\ Letters
              {\bf 73}, 2662 (1994);
            D0 collaboration: S.\ Abachi {\it et al.}, Phys.\ Rev.\ Letters
              {\bf 77}, 5011 (1996).
\bibitem{GSN} W.\ Tittel, J.\ Brendel, B.\ Gisin, T.\ Herzog, H.\ Zbinden,
              and N.\ Gisin, preprint quant-ph/9707042 (1997).
\bibitem{SC} E.\  Schr\"{o}dinger, Naturwissenschaften {\bf 23}, 807 (1935).
\bibitem{Ev} H.\ Everett, Rev.\ Mod.\ Phys.\ {\bf 29}, 454 (1957).
\bibitem{dB} L. de Broglie, J.\ Physique, 6e s\'{e}rie {\bf 8}, 225 (1927).
\bibitem{Bohm} D.\ Bohm, Phys.\ Rev.\ {\bf 85}, 166 and 180 (1952).
\bibitem{modal} B.\ van Fraassen, Synthese {\bf 42}, 155 (1979); S.\ Kochen,
      in {\it Symposium on the Foundations of Quantum Mechanics}, P.\ Lahti
      and P.\ Mittelstaedt, eds.\ (World Scientific, Singapore, 1985);
      R.\ A.\ Healey, {\it The Philosophy of Quantum Mechanics: An 
      Interactive Interpretation} (Cambridge University Press, 1989);
      D.\ Dieks, Phys. Letters A {\bf 142}, 439 (1989); J.\ Bub, Found.\
      Phys.\ {\bf 22}, 737 (1992).  
\bibitem{V} L.\  Vaidman, preprint quant-ph/9609006 (1996).
\bibitem{Al} D.\ Z.\ Albert, Phys.\ Letters {\bf 98A}, 249 (1983).
\bibitem{Eng} B.\ Englert, M.\ O.\ Scully, G.\ Sussman and H.\ Walther,
              Z.\ Naturforsch.\ {\bf 47a}, 1175 (1992).
\end{thebibliography}
\end{document}